\documentclass{article}
\title{Value of interparticle interaction potential as a
variable in solving many-body Schr\"{o}dinger equation}
\author{ V.M.Tapilin\\
  \it Boreskov Institute of Catalysis, Novosibirsk 630090, Russia}
\date{}
\begin{document}
\maketitle
\begin{abstract}
A many-body wave function is approximated by a product of two 
functions: the wave function $\phi$ depending on the particle 
coordinates and the 
function $\chi$ depending only on the value of interparticle interaction 
potential. For the given $\phi$ an ordinary linear differential equation 
for $\chi$ is derived by averaging the Hamiltonian over the constant  
interparticle interaction potential surface. Generalized Hartree-Fock
equations containing correlation effects are obtained. To test the 
proposed technique the ground $1^{1}S_0$ and excited $2^{3}S_1$ states 
of two-electron ions from H$^-$ up to Ne$^{8+}$ are calculated. In all 
cases the calculated energies are more accurate than those obtained with 
the Hartree-Fock theory even taking as $\phi$ the symmetrized product of 
electron wave functions in the Coulomb field of nucleus complitly 
disregarded the electron-electron interaction. Variation of factors in 
the one-particle wave function exponents leads to the results close to 
those of configuration interaction approach.
\end{abstract}

\vspace{1mm}
PACS: 31.15.-p, 31.15.Ar, 31.25.Eb, 31.25.Jf

\begin{center}
I. INTRODUCTION
\end{center}
Direct solution of the many-body Schr\"{o}dinger equation has been
performed only for two electron ions [1,2]. For more complicated cases
dimensionality of the equation and impossibility to separate 
variables in it force to search for approximating solutions. There are 
two main approaches to find these solutions: the Hartree-Fock 
approximation 
(HF) [3,4], and the density functional theory (DFT) [5,6]. The first one 
is based on the approximation  which considers electron moving in an 
average field of other electrons. The wave function in this 
approximation is presented by a symmetrized product of one particle 
functions, so the probability to find an electron at some space point 
does not depend on the positions of the rest electrons. To improve the 
Hartre-Fock theory one has to introduce this dependence known as the
correlation in electron movement [7]. It can be achieved through 
configuration interactions, perturbations or coupled cluster theory [8]. 
DFT rests on the density functional theory where exchange and 
correlation are introduced by a functional of electron charge density. 
Unfortunately, the exact form of the functional is not known and 
some approximations to it are used. In both approaches one needs 
to solve non-linear one-particle equations using an iteration procedure 
of the self-consistent field. In resent years, powerful computers and 
sophisticated computational algorithms have made a remarkable progress 
in this area. Nevertheless, the final solution of the problem has not 
been achieved yet, and developing new approaches, permitting to view 
this problem from another perspective, is being continued [9]. Searching 
for highly precise functions containing thousands variational parameters
as well as simple but accurate functions for heliumlike ions has not 
been stop yet also [10-25].  

The aim of the present paper is to propose a new approach to solve the 
many-body Schr\"{o}dinger equation going beyond HF theory.
The idea of this work has been stated in [26]. We consider the value 
of the interparticle interaction potential as a variable and represent 
a many-body wave function as a product of a function $\phi$ depending 
on particle coordinates and a function $\chi$ depending only on the 
value of interpaticle interaction potential. Function $\chi$ due to its 
dependence on this potential introduces correlation in the many-body 
wave function omitted in $\phi$. For a given $\phi$, using averaging the 
many-particle Hamiltonian over constant interaction potential 
hypersurface, the Schr\"{o}dinger equation is  transformed into a 
one-dimensional linear differential equation for $\chi$. In its turn, 
for any $\chi$ one can find the best $\phi$ which minimizes the total 
energy of the many-body system. Below, we develop the correspondent 
equations and apply them to calculate energies of two-electron ions.
  
\begin{center}
     II. EQUATIONS
\end{center}
The Schr\"{o}dinger equation of interacting particles can be written
in the form
\begin{equation} \label{schr}
[H_{0}(\mathbf{r})+v_{int}(\mathbf{r})]\Psi(\mathbf{r})=
E\Psi(\mathbf{r}), 
\end{equation}
where $H_{0}$ and $v_{int}$ are contributions to the Hamiltonian
from non-interacting particles and the interparticle interaction,
correspondently,
\begin{equation} \label{H1}
 H_{0}(\mathbf{r})=-\frac{1}{2}\sum_{i}\nabla^{2}_{i}+V(\mathbf{r}),
\end{equation} 
$\mathbf{r}$ is a vector with components $\mathbf{r}_{i}$, $i$ 
numerates particles, $V$ is a potential of an external field. 
For the Coulomb interaction
\begin{equation} \label{H2}
 v_{int}(\mathbf{r})=\frac{1}{p(\mathbf{r})}=
\sum_{i=1}^{n-1}\sum_{j>i}^{n}\frac{1}{r_{ij}}, 
\end{equation} 
where $r_{ij}=|\mathbf{r}_{i}-\mathbf{r}_{j}|$.

Let us introduce a function of the form
\begin{equation}\label{auxf}
\psi(\mathbf{r})=\phi(\mathbf{r})\chi(p).
\end{equation}
The function $\phi$ is supposed to be symmetrized in respect to the 
particle permutations. The type of symmetrization must reflect the 
total spin of the considering system [27]. Substitution of (\ref{auxf})
in (\ref{schr}) leads to
\begin{eqnarray}\label{eqnc}
-\frac{1}{2}\phi(\mathbf{r})\sum_{i}\Big(\nabla_{i}p\Big)^{2}
\frac{d^{2}\chi(p)}{d p^{2}}-
\Big[\frac{1}{2}\phi(\mathbf{r})\sum_{i}\nabla_{i}^{2}p+
\sum_{i}\nabla_{i} \phi(\mathbf{r})\nabla_{i}p\Big]
\frac{d\chi(p)}{d p}        \nonumber \\
+(h(p)+1/p)\phi(\mathbf{r})\chi(p)=E\phi(\mathbf{r})\chi(p)
\end{eqnarray}
It is easy to find
\begin{eqnarray}\label{grad1}
\nabla_{i}p=-\frac{1}{v_{int}^2}\nabla_{i}v_{int}=p^2\sum_{j\ne i}
\frac{\mathbf{r}_{ij}}{r_{ij}^3}\\ \label{grad2}
\nabla_{i}^2 p=\frac{2}{v_{int}^3}(\nabla_{i}v_{int})^2-
\frac{1}{v_{int}^2}\nabla_{i}^2 v_{int}=2p^3
\Big(\sum_{j\ne i}\frac{\mathbf{r}_{ij}}{r_{ij}^3}\Big)^2
\end{eqnarray}
In (\ref{grad2}) it was taken into account that the Coulomb potential is
satisfied to the Laplace equation.

The demands for function $\chi$ to minimize the value of the functional
\begin{equation}\label{efunc}
E=\langle \psi|H_{0}+v_{int}|\psi\rangle/\langle\psi|\psi\rangle
\end{equation}
for a preset $\phi$ leads to the equation for $\chi$ 
\begin{equation}\label{eqn}
-\frac{1}{2}t(p)\frac{d^{2}\chi(p)}{d p^{2}}-
u(p)\frac{d\chi(p)}{d p}+(h(p)+1/p)\chi(p)=E\chi(p),
\end{equation}
where
\begin{eqnarray} \nonumber
t(p)&=&s^{-1}(p)\int_{S(p)} d\mathbf{r}\phi^{*}(\mathbf{r})
\phi(\mathbf{r})\sum_{i}\Big(\nabla_{i}p\Big)^{2}, \\ 
\nonumber u(p)&=&s^{-1}(p)\int_{S(p)} d\mathbf{r}\phi^{*}
(\mathbf{r})\Big[\frac{1}{2}\phi(\mathbf{r})\sum_{i}\nabla^{2}_{i}p+
\sum_{i}\nabla_{i} \phi(\mathbf{r})\nabla_{i}p\Big], \\ 
\nonumber h(p)&=&s^{-1}(p)\int_{S(p)} d\mathbf{r}\phi^{*}(\mathbf{r})
H_{0}\phi(\mathbf{r}),  \\
s(p)&=&\int_{S(p)} d\mathbf{r}\phi^{*}(\mathbf{r})
\phi(\mathbf{r}), \label{s}
\end{eqnarray}
and the integration is performed over a constant interaction potential 
surface. The boundary conditions for $\chi$ follow from the demand
for $\psi$ to be finite in the whole space. 

Eq. (\ref{eqn}) is a linear ordinary differential equation for $\chi$. 
The term $h(p)$ is the contribution of $H_{0}$ to the total energy when 
the system is in the state $\phi$. The first and the second terms 
describe additional contributions to kinetic energy which doesn't enter 
$h(p)$ and appears due to the interparticle interaction $1/p$. If the 
interaction is absent and $\phi$ is an eigenstate of $H_{0}$ with energy 
$\epsilon$, the additional contribution must be zero that occur when
$\chi\equiv const$, $h$ is independent of $p$ and equals to $\epsilon$,
and $E=\epsilon$. For interacting particles, $\chi$ can be
a constant only if $\phi$ would be an exact solution of the 
Schr\"{o}dinger equation. The function $\chi$ we will name the correction 
function. The type of correction in $\chi$ depends on $\phi$. If $\phi$
corresponds to non-interacting particles $\chi$ contains corrections
due to the whole interaction, in case of HF $\phi$ only correlation 
part of the interaction, in the case of CI the part of correlation not 
taken into account in the used CI approximation.

Switching on the interaction between particles located on the same
constant interaction potential surface does not change the particle 
motion on this surface and can accelerate the particles only along the 
normal to the surface. Eq. (\ref{eqn}) takes into account this 
acceleration. However, the normal to the surface at point 
$\mathbf{r}$ can not be parallel to the normal at point $\mathbf{r}'$, 
so the acceleration obtained by the particle at $\mathbf{r}$ can have a 
tangential component at $\mathbf{r}'$. Eq. (\ref{eqn}) disregards this
effect and relative value in the change of the electron movement in 
normal and tangential directions due to electron-electron interaction 
for a given $\chi$ will determine the accuracy of the proposed 
approximation. To improve the approximation one can use common 
techniques choosing the better $\phi$.

The best $\phi$ for independent particle approximation can be obtained 
by minimization of functional 
(\ref{efunc}) considering $\chi$ as a known function. Presenting 
$\phi$ by a determinant of one-particle functions $\varphi_{i}$, one can 
obtain a set of equations analogous to the Hartree-Fock ones
\begin{eqnarray}\label{hf}
[-\frac{1}{2}\nabla_{1}^{2}+V(\mathbf{r}_{1})]
\varphi_{i}(\mathbf{r}_{1})+\sum_{jk}\tau_{ik}^{-1}[u_{kj}
(\mathbf{r}_{1})
\nabla_{1}\varphi_{i}(\mathbf{r}_{1})+ \nonumber \\
\upsilon_{kj}^{eff}(\mathbf{r}_{1})\varphi_{j}(\mathbf{r}_{1})]
=E\varphi_{i}(\mathbf{r}_{1}),
\end{eqnarray}
where
\begin{eqnarray} \nonumber
\upsilon_{ij}^{eff}(\mathbf{r}_{1})=
\langle\phi_{i}|\frac{\chi^{*}\chi}{p}+\sum_{k}\chi^{*}[
(\nabla_{k}p)^{2}\frac{d^{2}\chi}{dp^{2}}+
\nabla^{2}_{k}p\frac{d\chi}{dp}]|\phi_{j}\rangle_{1}, \\
\tau_{ij}(\mathbf{r}_{1})=\langle\phi_{i}|\chi^{*}\chi
|\phi_{j}\rangle_{1}, \label{tau}, \hspace{2mm}
u_{ij}(\mathbf{r}_{1})=
\langle\phi_{i}|\sum_{k}\chi^{*}\nabla_{k}p\frac{d\chi}{dp}
|\phi_{j}\rangle_{1},\label{ueff} 
\end{eqnarray}
and $\tau^{-1}$ is the reverse matrix to $\tau$.
Integration in (\ref{ueff}) is performed over coordinates 
of all particles except particle 1, $\phi_{i}$ means that function
$\varphi_{i}$ in $\phi$ is replaced by unity.
The set of equation (\ref{hf}) resembles the Hartree-Fock equations. A 
difference arises from the fact that minimization in respect to 
one-particle functions is performed with weighted function $\chi$ 
depending on the coordinates of all particles. It leads to the 
appearance of non-diagonal matrix elements in $\tau$ and replacing
the interpaticle Coulomb interaction $1/r_{12}$ of the Hartree-Fock 
equations by the screened interaction $\upsilon^{eff}$. Through this
interaction the correlation in the Hartree-Fock equations is introduced. 
For $\chi\equiv 1$, equations (\ref{hf}) are common Hartree-Fock 
equations.

Another possibility to improve $\phi$ avoiding the need to solve
HF equation occurs for $\phi$ depending on some parameters. In this
case eigenvalues of (\ref{eqn}) can be considered as the functional on
these parameters and their values can be find minimizing the 
functional. This technique will be used in section 3.

Constructing interaction potential surface, it is useful to note that 
the form of the interparticle interaction (\ref{H2}) allows to carry any 
set of particle positions $\mathbf{r}$ belonging to constant interpaticle 
interaction potential surface $1/p$ to another set $\mathbf{r}'$ 
belonging to surface $1/p'$ by coordinate scaling with factor $p'/p$. It 
makes the generating points for any surface an easy task and restricts
itself for initial generation by the surface, for example, $p=1$. To 
obtain the point for any other surface one can use coordinate scaling 
again. Besides, the construction of the constant interaction potential 
surface can be presented in a recurrent form. For this purpose let us 
introduce notations 
\begin{equation}\label{Qeq}
\frac{1}{q_{n}}=\sum_{i=1}^{n-1}\frac{1}{r_{in}},
\end{equation}
where $n$ is the number of particles, and express the interaction
potential as contributions from $n-1$ particles and interaction of the
$n$th particle with the rest
\begin{equation}\label{Q}
\frac{1}{p_{n}}=\frac{1}{p_{n-1}}+\frac{1}{q_{n}}.
\end{equation}
Then integration over the constant interaction potential surface 
$S(p_{n})$ for n particles can be presented in the form
\begin{equation}\label{SN}
\int_{S(p_{n})}d\mathbf{R}_{n}=\int_{S(p_{n-1})}
d\mathbf{R}_{n-1}\int_{S(q_{n})}d\mathbf{r}_{n}
\end{equation}
where $ d\mathbf{R}_{n}=d\mathbf{r}_{1}\ldots d\mathbf{r}_{n}$. For a 
given $\mathbf{R}_{n-1}\equiv\{\mathbf{r}_{1},\ldots,\mathbf{r}_{n-1}\}$
integration over $\mathbf{r}_{n}$ is performed over the surface 
determined in 3-dimension space by (\ref{Q}). For two particles it is 
the sphere of radius $p$ around position of the first particle, for 
three particles it is a circle formed by crossing the sphere of radius
$r_{13}$ around $\mathbf{r}_1$ with the sphere of radius $r_{23}$ around
$\mathbf{r}_2$, for four particles it is only two points at the circle,
etc. In addition one can restrict itself by irreducible part of the 
surface results from the identity of the particles. Evidently, with the 
help of particle numeration one can always satisfy the condition
\begin{equation}\label{cond}
\sum_{i}^{n}\frac{1}{r_{il}}\le\sum_{i}^{n}\frac{1}{r_{ik}}\hspace{3mm}
\mbox{if}\hspace{3mm}l<k,
\end{equation}
and it will be enough to consider only the part of the whole surface,
satisfying to (\ref{cond}).

As an example, we represent the integral over the constant interaction 
potential surface for two particles 
\begin{eqnarray}\label{srf2} \nonumber
\int_{S(p)}d\mathbf{R}_{2}&=&\int_{0}^{\infty}r_{1}^{2}d\mathbf{r}
_{1}
\int_{0}^{2\pi}d\varphi_{1}\int_{0}^{\pi}\sin{\theta_{1}}d\theta_{1}\\
&&\int_{0}^{\infty}\delta(r_{12}-p)r_{12}^{2}dr_{12}
\int_{0}^{2\pi}d\varphi_{2}\int_{0}^{\pi}\sin{\theta_{2}}d\theta_{2}
\label{S2}, \end{eqnarray}
where the integration over the coordinates of the 1st particle is performed
over the whole space, while the integration over the coordinates of the
2nd particle for a given $\mathbf{r}_{1}$ is reduced to the integration 
over a sphere of radius $r_{12}=p$ around the first particle. 

\begin{center}
III. HELIUMLIKE IONS
\end{center}
As an application of the developed technique we calculated the ground 
$1^{1}S_{0}$ and excited $2^{3}S_{1}$ states electronic structure
of heliumlike ions from H$^{-}$ up to Ne$^{8+}$. For these ions 
$p\equiv r_{12}$ and (\ref{auxf}) is reduced to the form used in a 
number of papers [1,2,17,22,23,25]. The difference between those and the present 
paper consists in the different ways of the wave function calculation.

We represent
\begin{equation}\label{phi2}
\phi(\mathbf{r}_{1},\mathbf{r}_{2})=[\varphi_1(r_1)\varphi_2(r_2)\pm
\varphi_2(r_1)\varphi_1(r_2)]/\sqrt{2(1\pm S^2)}
\end{equation}
where $S=\langle \varphi_1|\varphi_2\rangle$, for $1^{1}S_{0}$
\begin{equation}\label{f1s}
\varphi_{i}(r)=2Z_i^{3/2}\exp{(-Z_{i} r)}
\end{equation}
is $1s$ function of the electron in the field of nucleus with the charge 
$Z_i$. For $2^{3}S_{1}$ the function of type (\ref{f1s}) 
was used for $\varphi_1$, whereas
\begin{equation}\label{f2s}
\varphi_{2}(r)=Z_2^{3/2}(1-Z_2 r/2)\exp{(-Z_2 r/2)}/
\sqrt{2}
\end{equation}
is the $2s$ function of the electron in the field of nucleus with the 
charge $Z_2$. Factor $Z_i$ in function (\ref{f1s}) or (\ref{f2s}) is 
known as an efficient nuclear charge seeing by the electron in the 
corresponding state, $Z$ without subscript we save for the real nuclear 
charge. For function (\ref{phi2}) Eq. (\ref{eqnc}) can be written in 
the form
\begin{eqnarray}\label{eqn1} \nonumber
-\phi\frac{d^{2}\chi}{dp^2}-\Big[\frac{p^2+r_1^2-r_2^2}{2pr_1}
\frac{\partial\phi}{\partial r_1}+
\frac{p^2+r_2^2-r_1^2}{2pr_2}\frac{\partial\phi}{\partial r_1}+
\frac{2}{p}\phi\Big]\frac{d\chi}{dp}\\
-\Big[\frac{1}{2}\frac{\partial^2\phi}{\partial r_1^2}
+\frac{1}{2}\frac{\partial^2\phi}{\partial r_2^2} 
+\frac{1}{r_1}\frac{\partial\phi}{\partial r_1}
+\frac{1}{r_2}\frac{\partial\phi}{\partial r_2}+\frac{1}{r_1}
+\frac{1}{r_2}-\frac{1}{Zp}\Big]\chi=E\phi\chi
\end{eqnarray}
Here $Z$ is the nuclear charge, the length unit is Bohr/Z, and atomic
units are used for other values. 

Integration in (\ref{srf2}) over a constant potential surface, when 
calculating expressions (\ref{s}) for these case,  is reduced to 
\begin{equation}
\int_{S(p)}d\mathbf{R}_{2}=4\pi p\int_{0}^{\infty}r_1dr_1\int_{|r_1-p|}
^{r_1+p}r_2dr_2
\end{equation}
and can be 
performed analytically. The results are cumbersome and we present only
their common form
\begin{equation}\label{intgf}
\sum_{k}c_{1,k} p^k e^{-2Z_1 p}+
\sum_{l}c_{2,l} p^l e^{-(Z_1+Z_2) p}+
\sum_{m}c_{3,m} p^m e^{-2Z_2 p} 
\end{equation}
Eq. (\ref{eqn}) with coefficients (\ref{intgf}) can be solved 
analytically for asymptotic cases $p\to 0$ and $p\to \infty$. The 
finite asymptotic solutions are
\begin{eqnarray}\label{as1}
\lim\limits_{p\to 0}\chi(p)\to\chi(0)(1+\frac{p}{2Z}+\frac{p^2}{12Z^2}+
O(p^3)),\\   \label{as2}
\lim\limits_{p\to \infty}\chi(p)\to \exp{\big[\big(-Z_{min}-
\sqrt{Z_{min}^2-E+E_0}\big)p\big]}, 
\end{eqnarray}
where $Z_{min}=\min(Z_1,Z_2)$, $E_0$ is energy in state $\phi$. 

Eqs. (\ref{as1}) and (\ref{as2}) were taken as the bounder conditions 
in the numerical solution of (\ref{eqn}). For the numerical calculations 
the differential operators were approximated with finite differences.
Adopting
\begin{eqnarray}
\frac{d\chi_{i}}{dp}\approx\frac{\chi_{i+1}-\chi_{i-1}}{2\Delta p},  \\
\frac{d^{2}\chi_{i}}{dp^2}\approx\frac{\chi_{i-1}+\chi_{i+1}-2\chi_{i}}
{\Delta p^2},
\end{eqnarray}
where $i$ numerates points of discrete set $\{p_{i}\}$, $\chi_{i}=
\chi(p_{i})$, $\Delta p=p_{i}-p_{i-1}$, Eq. (\ref{eqn}) can be 
represented by $n\times n$ tridiagonal matrix with $n$ is the number of 
points in $\{p_{i}\}$
\begin{eqnarray}
H_{ii}=\frac{t_i}{2\Delta p^2}+\frac{1}{Zp_i}+h_i, \\
H_{i,i\pm 1}=-\frac{t_i}{2\Delta p^2}\pm\frac{u_i}{2\Delta p} 
\end{eqnarray}
for $1<i<n$, and
\begin{eqnarray}\label{bnd1}
H_{11}=\frac{t_1}{2\Delta p^2}+\frac{1}{Zp_1}+h_i+\Big(-\frac{t_1}{2\Delta p^2}
+\frac{u_1}{2\Delta p}\Big)/(1+\Delta p/2Z+\Delta p^2/12Z^2), \\ \label{bnd2}
H_{n,n}=\Big(-\frac{t_n}{2\Delta p^2}+\frac{u_n}{2\Delta p}\Big)
\exp{\big[\big(-Z_{min}-\sqrt{Z_{min}^2-E+E_0}\big)\Delta p\big]} 
\end{eqnarray}
in accordance with (\ref{as1}), (\ref{as2}). Because boundary condition
(\ref{bnd2}) depends on calculating eigenvalue, it is satisfied in the 
iteration process. Eigenvalues and eigenfunctions of (\ref{eqn}) were 
approximated by eigenvalues and eigenvectors of matrix $H$ in interval 
$0<p<20/Z$ with 200 points at the interval. The value of interval and 
the number of points guarantee all significant numbers in the results 
presented below.

For $1^1S_0$ states the calculations were performed for three sets of 
function (\ref{f1s}): a) $Z_1=Z_2=Z$; b) $Z_1=Z_2$, however, the value 
is chosen from energy minimum; c) $Z_1\ne Z_2$ and each the value is 
chosen from energy minimum. For $2^3S_1$ only the first and the third 
cases were considered. The calculated energies of these states together 
with those of HF theory for $1^1S_0$, and configuration interaction 
for $1^1S_0$ and $2^3S_1$ are shown in Table I.

As it is seen in Table I in all cases the energies of $1^1S_0$ states 
calculated with Eq. (\ref{eqn}) are less than the energies calculated
with HF theory. Even for functions $\phi$ completely disregard the 
interaction between electrons the correction function $\chi$ capable to 
take into account electron-electron interaction more accurate than it is 
done in HF approximation. However, the achieved accuracy is still far 
from the accuracy obtained by configuration interaction calculations 
[29] or calculations which explicitly introduce the distance between 
electrons in the wave function [1,2,23,25]. It is due to correction 
function $\chi$ describe the electron motion only between different 
electron-electron potential surface keeping the motion on a surface 
determined by function $\phi$. 

The results can be improved by choosing the best $\phi$ solving Eq. 
(\ref{hf}). However, we attempt to use another way. Function (\ref{phi2}) 
depends on two parameters, $Z_1$ and $Z_2$, so eigenvalues of 
(\ref{eqn}) are the functionals of these parameters. The best 
one-particle functions of the form (\ref{f1s}) and (\ref{f2s}) minimize 
the functional. It permits us to avoid the solution of non-linear HF 
equations and makes technique based on (\ref{eqn}) self-sufficient. 
Table I contains obtained results. We observe a slight improvement when 
both electrons are seeing an equal nucleus electric charge. Considerable 
improvement is observed when different electrons are seeing different 
effective nucleus charges. These charges and their dependence on Z are 
presented in Fig. 1. 

As for $1^1S_0$ the results for $2^3S_1$ obtained with $Z_1=Z_2=Z$
can be significantly improved by introducing different effective nuclear 
charges for the electrons. As it is seeing in Table I, in this case
the results are close to those, obtained with configuration interaction.
The charges and their dependence on Z are presented in Fig. 2. 

The function $\phi$ obtained by minimization of $E(Z_1,Z_2)$ can be 
considered as HF function. Thus, the correction function $\chi$ can be
regarded as correlation functions. These functions, normalized so that
integral $\chi(p)s(p)\chi(p)$ over interval $0\le p\le 20$ equals to 1, 
are presented in Fig.3. The functions reduce the probability of
electrons to be at a short distances showing monotonic growth with the
distance between the electrons. The behavior of $\chi$ for all ions is 
similar, however, the functions become smoother with the growth of Z and
tend to some limit. The deviation $\chi$ from the unity for $2^3S_1$ 
states is less then for $1^1S_0$ state suggesting a decrease of 
correlation effects. It is known to occur due to the exchange hole 
appearing for antisymmetric $\phi$. The function $\chi$ changes this 
hole, with this change being more significant at electron-electron 
distances $p\le 1/Z$ reducing further the small probability to find 
electrons at these distances. 

The function $s(p)$ determined in (\ref{s}) is the distribution function of 
$r_{12}$ for wave function $\phi$, and function $\chi s\chi$ is the 
distribution function for $\phi\chi$. These function for $1^1S_0$ states
for $H^-$ and $Ne^{8+}$ are shown in Fig. 4. Besides, Fig. 4 shows
the corresponding correlation holes determined as $s(\chi^2-1)$. In 
general, the form of the correlation holes is the same as in [30]. 
However, the depths of the holes decreases from $H^-$ to $Ne^{8+}$. The 
distribution functions and correlations holes for $2^3S_1$ of $He$ and 
$Ne^{8+}$ are shown in Fig. 5. As for $1^1S_0$ states the depths of the 
holes decreases with nuclear charge growth and the holes become more
shallow in comparison with $1^1S_0$.

Until now we have considered only the lowest eigenstates of (\ref{eqn}).
However, for given the $\phi$ we can calculate higher eigenstates for 
$\chi$. We will demonstrate how these states correspond to real excited 
states on the example of non-interacting electron for which the exact 
results have been available. In Fig.~6 functions $\chi$ for the lowest 
three solutions of (\ref{eqn}) for non-interacting electrons are shown. 
As it is seen, higher eigenfunctions exhibit nodes due to which 
function $\phi$ based on $1s$ functions is transformed in function 
$\phi$ similar to the function based on $2s$ or $3s$ functions. However, 
the energies corresponding to these function are $E=-0.440$ and $-0.245$ 
instead of exact energies -0.625 and -0.555 a.u. 
The source of pure accuracy is the conservation $1s$ function in 
computations (\ref{s}). The use of proper function in (\ref{s}) makes it
possible to obtain the exact results for the excited states. As such, we 
can use higher eigenstates of (\ref{eqn}) only for a crude estimation 
of exited states energies and use proper $\phi$ to obtain more 
accurate results.

\begin{center}
{IV. CONCLUSIONS}
\end{center}
Eq. (\ref{eqn}) reduces the many-body Schr\"{o}dinger equation to the
2nd order linear ordinary differential equation for the correction 
function. It describes the changes in electron movement between constant
interaction potential surfaces due to electron-electron interaction. The
changes in electron movement along the surface are disregarded in (\ref{eqn}).

Solutions of the (\ref{eqn}) for two electron ions give the 
total energies more accurate than HF theory does. It is true even in the 
simplest case where wave functions of non-interacting electrons are used 
for function $\phi$. For $\phi$ containing only two variations 
parameters, efficient nuclear charges visible by the electrons, the 
results are close to those obtained by configuration interaction or 
explicit dependence of the wave function on the distance between 
electrons where tens parameters are used. It means that the
electron-electron interaction changes the electron motion of independent 
particle mainly between different constant interaction potential surfaces 
affecting the movement along the surfaces to a lesser degree. 
Nevertheless, to improve the accuracy it is needed, in one or another 
way, to take into account the changes in electron movement along the 
surfaces.

Considering eigenvalues of (\ref{eqn}) as a functional of parameters 
determining $\phi$, it seems possible to avoid the solution of 
non-linear HF equation when finding the best $\phi$ by solving 
iterative Eq. (\ref{eqn}). Hopefully, Eq. (\ref{eqn}) can also replace
the HF equations for system with more then two electrons. The main
effort in the proposed technique application is the integration 
over constant interaction potential surface. For a two electron system 
it can be done analytical. It doesn't seem difficult to solve this 
problem for a several electron system. However, computation efforts will 
grow with the increasing number of electrons and, eventually, the direct 
integration in (\ref{s}) will be inefficient. We will put off discussion the
possible ways to solve this problem after solving it for several 
electron systems.

\begin{center}
{ACKNOWLEDGMENT}
\end{center}
This work is supported by Interdisciplinary Project \# 74 of the 
Siberian Branch of the Russian Academy of Science.

\pagebreak

\noindent
1. E.A. Hylleraas, Zs. Phys. {\bf 65}, 209 (1930).

\noindent
2. C. Eckart, Phys. Rev. {\bf 36}, 878 (1930).

\noindent
3. D.R. Hartree, Proc. Camb. Phil. Soc.{\bf 24}, 89, 111, 426 
(1928).

\noindent
4. V. Fock, Zs. Phys. {\bf 61}, 126 (1930).

\noindent
5. P. Hohenberg and W. Kohn, Phys. Rev. B {\bf 136}, 864 (1964).

\noindent
6. W. Kohn and L.J. Sham, Phys. Rev. A {\bf 140}, 1133 (1965).

\noindent
7. E. Wigner, Phys. Rev. {\bf 46}, 1002 (1934).

\noindent
8. A. Szabo and N.S. Ostlund, {\it Modern Quantum Chemistry} 
(McGraw-Hill, New York, 1989).

\noindent
9. P.M.W. Gill, D.L. Crittenden, D.P. O'Neill and N.A. Besley,
Phys. Chem. Chem. Phys. {\bf 8}, 15 (2006).

\noindent
10. E.A. Hylleraas and J. Midtdal, Phys. Rev. {\bf 103}, 829 (1956).

\noindent
11. J. Linderberg, Phys. Rev. {\bf 121}, 816 (1961).

\noindent
12. Y. Accad, C.L. Pekeris and B. Schiff, Phys. Rev. A {\bf 4}, 516 
(1971)

\noindent
13. R.J. Tweed, J. Phys. B: Atom. Molec. Phys. {\bf 5}, 810 (1972).

\noindent
14. M.H. Chen, K.T. Cheng and W.R. Johnson, Phys. Rev. A {\bf 47}, 3692
    (1993)

\noindent
14. A.J. Thakkar and T. Koga, Phys. Rev. A {\bf 50}, 854 (1994).

\noindent
16. R.A. Bonham and D.A. Kohl, J. Chem. Phys. {\bf 45}, 2471 (1996).

\noindent
17. C. Le Sech, J. Phys. B.: At. Mol. Opt. Phys. {\bf 30}, L47 (1997).

\noindent
18. S.P. Goldman, Phys. Rev. A {\bf 57}, R677 (1998).

\noindent
19. A.M. Frolov and V.H. Smith, J. Phys. B.: At. Mol. Opt. Phys.
   {\bf 37}, 2917 (2004).

\noindent
20. V. Korobov, Phys. Rev. A {\bf 69}, 0545012 (2004).

\noindent
21. S.H. Patil, Eur. J. Phys. {\bf 25}, 91 (2004).

\noindent
22. D.M. Mitnik and J.E. Miraglia, J. Phys. B.: At. Mol. Opt. Phys.
    {\bf 38}, 3325 (2005).

\noindent
23. K.V. Rodriguez and G. Gasaneo, J. Phys. B: At. Mol. 
 Opt. Phys. {\bf 38}, L259 (2005).

\noindent
24. C. Schwartz, Int. J. Mod. Phys. E {\bf 15}, 877 (2006).

\noindent
25. K.V. Rodriguez, G. Gasaneo and D.M. Mitnik, J. Phys. B: At. Mol. 
 Opt. Phys. {\bf 40}, 3923 (2007).

\noindent
26. V.M. Tapilin, Zh. Struct. Khimii {\bf 49}, 409 (2008) [J. Struct. 
Chem., to be translated].

\noindent
27. L.D. Landau and L.M. Lifshits,{\it Quantum Mechanics} (Pergamon, 
London, 1977).

\noindent
28. E. Clementi, C. Roetti, Atomic data and Nuclear data Tables
 {\bf 14}, 177 (1974).

\noindent
29. http://cdfe.sinp.msu.ru/services/wftables/he\_s1.htm

\noindent
30. W.A. Lester, M. Krauss, Jour. Chem. Phys. {\bf 41}, 1407 (1964)

\pagebreak

\vspace{15mm}
TABLE I. Energies of Heliumlike ions (a.u.): HF - HF 
approximation; Z - Eq. (\ref{eqn}) with effective nuclear charges 
equal to $Z$, $Z_1=Z_2\ne Z$, $Z_1\ne Z_2\ne Z$; CI - configuration 
interaction.
\begin{center}
\begin{tabular}{cccccccc} \hline \hline
ion & & $1^{1}S_0$&  $2^{3}S_1$&ion & & $1^{1}S_0$& $2^{3}S_1$ \\ \hline
$H^{-}$& HF  &          &     &$C^{4+}$& HF&-32.36137$^a$& \\                  
       & Z   &-0.49843  &     &        &Z  &-32.38345&-21.40351 \\
       & $Z_1=Z_2$   &-0.50936  &     &      & $Z_1=Z_2$ &-32.39605& \\                  
  &$Z_1\ne Z_2$   &-0.52623  & & & $Z_1\ne Z_2$ &-32.40493&-21.41659                                          \\         
       & CI &-0.52760$^b$&     &  &CI&-32.40544$^b$&-21.42249$^c$ \\                           
He  &HF &-2.86171$^a$ &  &        $N^{5+}$ & HF  &-44.73618$^a$& \\                  
    &  Z&-2.87940 &-2.15491 & &       Z &-44.75873&-29.34061 \\
    & $Z_1=Z_2$&-2.89142 &  &             & $Z_1=Z_2$ &-44.77422& \\                  
 &$Z_1\ne Z_2$&-2.90208 &-2.17087 & &$Z_1\ne Z_2$ &-44.78001&-29.35355\\         
 &CI&-2.90325$^b$ &-2.17521$^c$  & &CI&-44.78061$^b$&-29.36118$^c$ \\                           
$Li^{+}$ &HF &-7.23633$^a$ &   &  $O^{6+}$ & HF&-59.11159$^a$& \\                  
         & Z &-7.25642 &-5.09213 &     & Z   &-59.13392&-38.52770 \\
         & $Z_1=Z_2$ &-7.26856 & &        & $Z_1=Z_2$   &-59.15135& \\                  
 &$Z_1\ne Z_2$&-7.27829 &-5.10636 &&$Z_1\ne Z_2$&-59.15543&-38.54054 \\         
   &CI&-7.27928$^b$ &-5.11075$^c$ & &CI&-59.15574$^b$&-38.55098$^c$ \\                           
$Be^{2+}$& HF&-13.61130$^a$&      & $F^{7+}$ & HF&-75.48702$^a$& \\                   
         & Z &-13.63244&-9.27928&&       Z   &-75.50906&-48.96479 \\   
         & $Z_1=Z_2$ &-13.64393& &    &  $Z_1=Z_2$   &-75.52834& \\
 &$Z_1\ne Z_2$&-13.65406&-9.29290 &&$Z_1\ne Z_2$&-75.53066&-48.97754 \\          
   &CI&-13.65485$^b$&-9.29739$^c$ &  &CI&-75.53083$^b$&-48.99223$^c$ \\                            
$B^{3+}$ & HF&-21.98607$^a$& &      $Ne^{8+}$& HF&-93.86174$^a$& \\                  
         & Z &-22.00805&-14.71640& &       Z &-93.88415&-60.65187 \\
         & $Z_1=Z_2$ &-22.02112& &         &  $Z_1=Z_2$ &-93.90520& \\         
 &$Z_1\ne Z_2$&-22.02934&-14.72969& &$Z_1\ne Z_2$&-93.90577&-60.66457\\                  
  &CI&-22.03020$^b$&-14.73463$^c$& &CI&-93.90592$^b$&-60.68527$^c$\\ 
\hline \hline             

$^a$Ref. [28]  \\
$^b$Ref. [29]  \\
$^c$Ref. [12]  \\
\end{tabular}
\end{center}

\pagebreak
\noindent
FIG 1. Efficient charges for $1^1S_0$ states.\\
FIG 2. Efficient charges for $2^3S_1$ states.\\
FIG 3. Functions $\chi$ for $1^1S_0$ and $2^3S_1$ states. \\
FIG 4. Distribution of the electron-electron distance (solid and dash
       lines for the HF approximation and present calculation, 
       correspondently) and correlation holes for $1^1S_0$ states. \\
FIG 5. Distribution of the electron-electron distance (solid and dash
       lines for the HF approximation and present calculation, 
       correspondently) and correlation holes for $2^3S_1$ states. \\
FIG 6. Function $\chi$ of the ground state and excited states for 
       non-interacting particles.

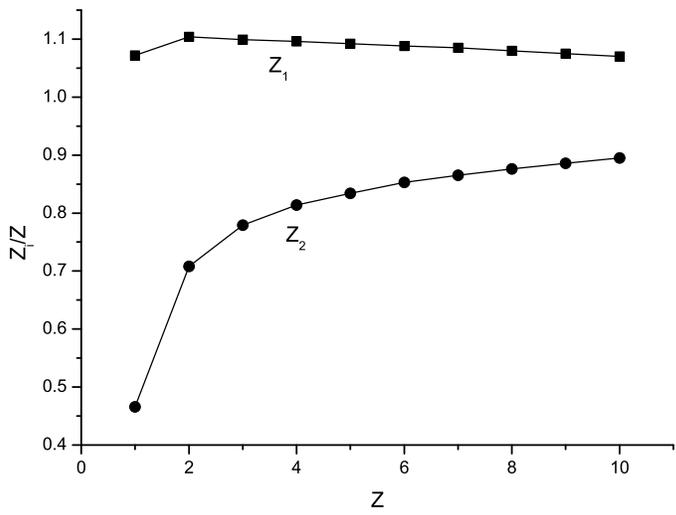
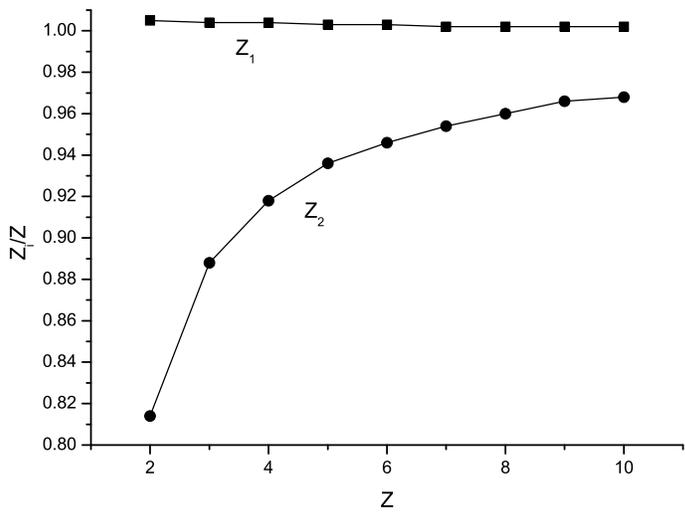
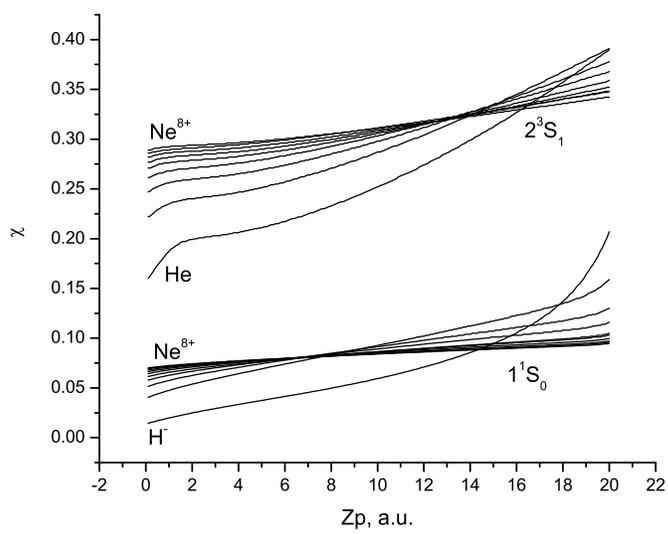
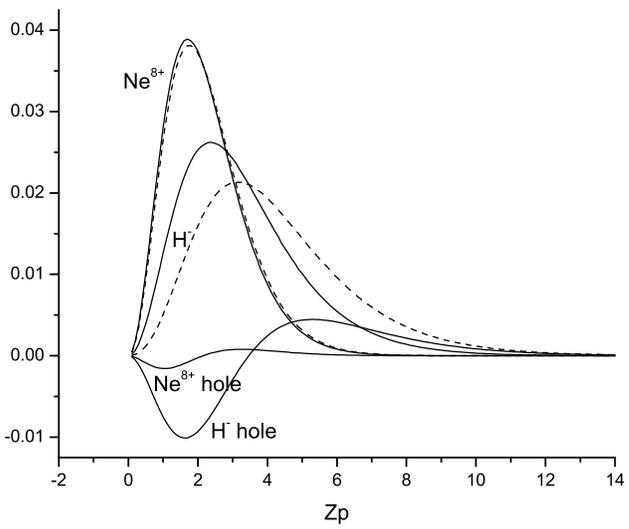
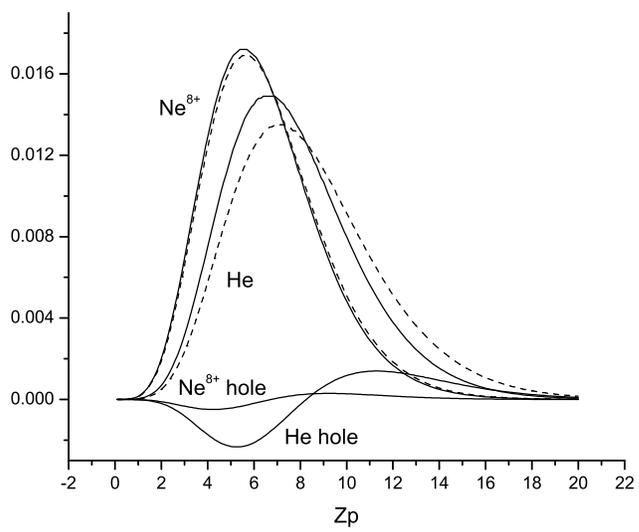
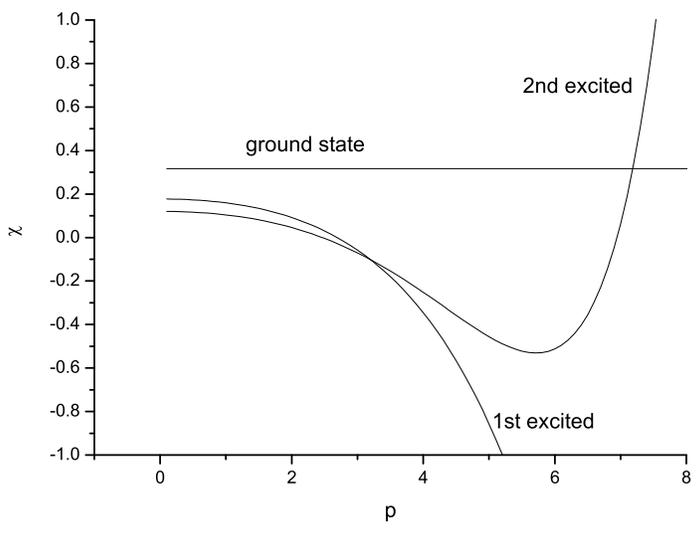
\end{document}